%% file: main.tex
\documentclass[10pt,conference]{IEEEtran}
\IEEEoverridecommandlockouts

\usepackage{float}
\usepackage{cite}
\usepackage{amsmath,amssymb,amsfonts}
\usepackage{graphicx}
\usepackage{bm}
\usepackage{booktabs} 
\usepackage{diagbox} 
\usepackage{enumitem}
\usepackage{multicol}
\usepackage{multirow}
\usepackage{makecell}
\usepackage{pifont}
\usepackage{subfigure}
\usepackage{xcolor}
\usepackage[linesnumbered,lined,ruled,commentsnumbered]{algorithm2e}
\usepackage{url,hyperref,cleveref}

\def\BibTeX{{\rm B\kern-.05em{\sc i\kern-.025em b}\kern-.08em
    T\kern-.1667em\lower.7ex\hbox{E}\kern-.125emX}}

\begin{document}

\title{Masked Self-distilled Transducer-based Keyword Spotting with Semi-autoregressive Decoding}

\author{\IEEEauthorblockN{Yu Xi$^{1,2}$, Xiaoyu Gu$^1$, Haoyu Li$^2$, Jun Song$^2$,  Bo Zheng$^2$, Kai Yu$^{1{\dagger}}$ \thanks {$^{\dagger}$ Corresponding Author.}}

\IEEEauthorblockA{\textit{$^1$MoE Key Lab of Artificial Intelligence, AI Institute, X-LANCE Lab, Shanghai Jiao Tong University, Shanghai, China}}
\IEEEauthorblockA{\textit{$^2$Taobao \& Tmall Group of Alibaba, China}}
\IEEEauthorblockA{
\{yuxi.cs, kai.yu\}@sjtu.edu.cn~~~ \{jsong.sj, bozheng\}@alibaba-inc.com
}}


\maketitle

\begin{abstract}
RNN-T-based keyword spotting (KWS) with autoregressive decoding~(AR) has gained attention due to its streaming architecture and superior performance. However, the simplicity of the prediction network in RNN-T poses an overfitting issue, especially under challenging scenarios, resulting in degraded performance. In this paper, we propose a masked self-distillation (MSD) training strategy that avoids RNN-Ts overly relying on prediction networks to alleviate overfitting. Such training enables masked non-autoregressive (NAR) decoding, which fully masks the RNN-T predictor output during KWS decoding. In addition, we propose a semi-autoregressive (SAR) decoding approach to integrate the advantages of AR and NAR decoding. Our experiments across multiple KWS datasets demonstrate that MSD training effectively alleviates overfitting. The SAR decoding method preserves the superior performance of AR decoding while benefits from the overfitting suppression of NAR decoding, achieving excellent results.
\end{abstract}


\section{Introduction}


Keyword Spotting (KWS) \cite{paper-guoguochen-dnn-kws, hmm-filler-1, hmm-filler-2, icassp2022_yuxi_text-adap,li24r_interspeech,taslp2025-yuxi-mfakws} is designed to detect predefined keywords in continuous speech, with a keyword-based interaction front-end being crucial for smart devices. Due to constraints in computation and memory, designing compact yet powerful KWS models has become a challenging research topic. With the natural streaming capability and superior performance, RNN-Ts have achieved great success in multiple research areas, such as automatic speech recognition~(ASR) \cite{RNNT-ASR_01, RNNT-ASR_02,RNNT-ASR_03, RNNT-ASR_04, RNNT-ASR_05}, speech translation~(ST)~\cite{RNNT-ST_01}, and text-to-speech~(TTS)~\cite{arxiv2024_chenpengdu_vallt}. Transducer-based KWS systems also achieves significant attention. \cite{RNNT-KWS_01} introduces a bias module for RNN-T to enhance keyword detection. The authors in~\cite{RNNT-KWS_02,RNNT-KWS_03-RNNTByteDance} leverage extensive synthetic and real speech data to achieve high performance in RNN-T KWS. Subsequently,~\cite{RNNT-KWS_04,RNNT-KWS_05_CaTT-KWS} adopt multi-stage strategies to reduce false alarms. TDT-KWS~\cite{RNNT-KWS_06_TDT} proposes an efficient streaming decoding algorithm for RNN-T KWS, utilizing a Token-and-Duration Transducer variant to accelerate inference speed.

In keyword spotting, positive training data typically consists solely of keyword sequences, resulting in fixed input transcripts and prediction targets for the prediction network. When large-scale keyword-specific data is introduced to enhance robustness, the prediction network may overfit to the keyword pattern, diminishing the model’s ability to discriminate between positive and negative samples and increasing false alarm rates. This issue is particularly pronounced in RNN-T-based KWS systems due to the simplicity of the prediction network architecture~\cite{RNNT-KWS_03-RNNTByteDance}.
Previous works have shown that using hybrid encoders or multi-stage strategies can effectively alleviate overfitting and reduce false alarms. 
For instance, \cite{RNNT-KWS_04} proposes a multi-level detection paradigm based on the posterior probabilities output by the Transducer acoustic model, which incrementally reduces false alarms. CaTT-KWS~\cite{RNNT-KWS_05_CaTT-KWS} integrates a Transducer-based keyword detector, a frame-level force alignment module, and a transformer-based decoder to create a multi-stage decoding process. U2-KWS~\cite{asru2023-aozhang-u2kws} uses the CTC branch and decoder branch as the first and second-stage models, respectively, to enhance detection reliability.

Although hybrid systems and multi-stage strategies mitigate overfitting, they introduce complexity into both the training procedure and decoding pipeline. Excessive manual operations and prior knowledge across multiple stages may lead to suboptimal performance for keyword detection. In this paper, we aim to identify a simpler yet more effective approach for Transducer-based models. Recently, a hybrid-autoregressive Transducer~\cite{xu2025threeinone} was introduced, which transforms Transducers into a non-autoregressive mode. The model learns to process speech without relying on the prediction network by randomly setting its output to zero during training. Inspired by this, we propose a self-distilled Transducer-based KWS model with a semi-autoregressive decoding strategy. 
Our core contributions can be summarized as follows:
\begin{itemize}
    \item We propose a masked self-distillation (MSD) training framework for on-device Transducer-based KWS systems. The MSD strategy enhances masked learning, enabling effective non-autoregressive (NAR) decoding without the prediction network.

    \item We propose a semi-autoregressive (SAR) decoding strategy that combines the high accuracy of standard AR decoding with the robustness and overfitting resistance of mask-based NAR decoding in noisy conditions. The SAR decoding algorithm can works well across different application scenarios.
    
    \item Experimental results show that our training and inference framework achieves strong performance on standard datasets and demonstrates improved robustness under challenging acoustic conditions.
\end{itemize}


\begin{figure*}[ht]
    \centering
    \includegraphics[width=\linewidth]{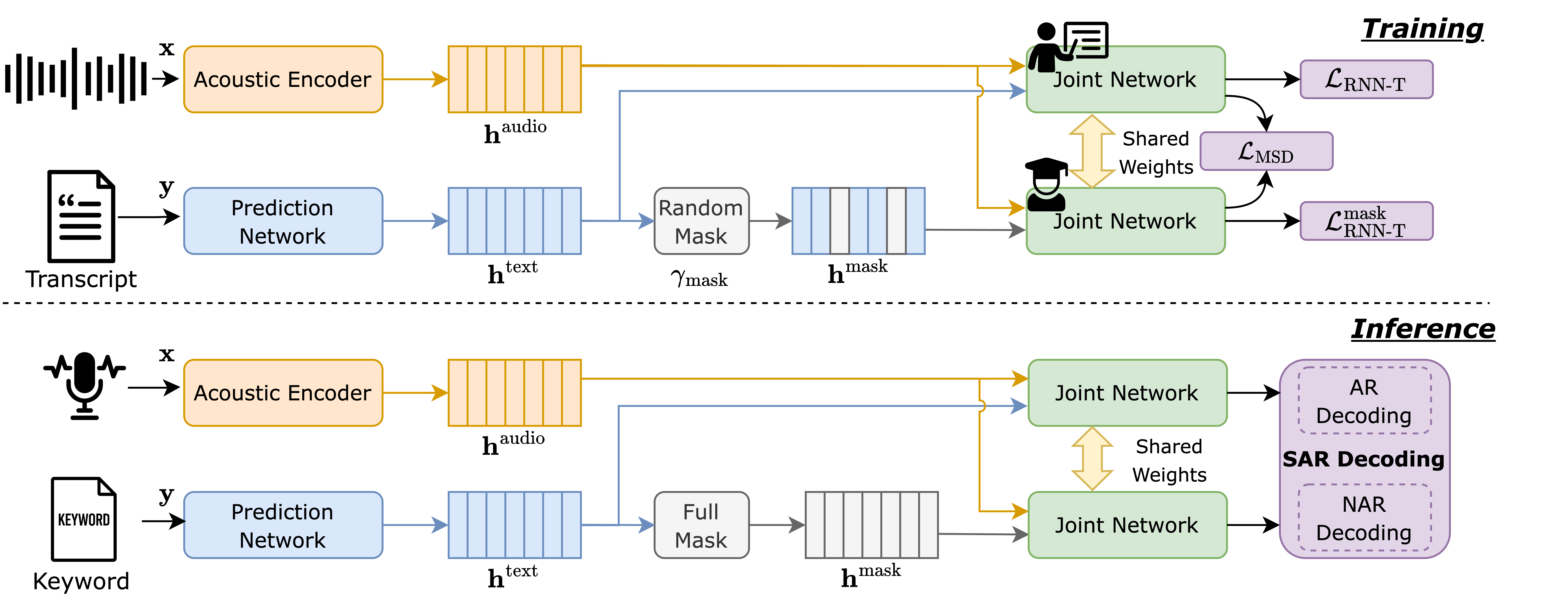}
    \caption{The training and inference overview of the proposed system. Both conventional and masked RNN-T training share the same architecture and parameters. During training, the prediction network is fed transcriptions, while it is fed keywords during inference.}
    \label{fig:overview}
\end{figure*}

\section{Preliminary: Transducer-based keyword spotting}
\label{sec:2.1}
Transducers consist of an encoder, a prediction network~(predictor), and a joint network~(joiner). The encoder transforms the acoustic features $\mathbf{x} = [x_{1}, x_{2}, ..., x_{T}] \in \mathbb{R}^{T \times F}$, where each $x_{t}$ is an acoustic feature vector with dimension $F$ and $T$ represents the total number of frames, into a latent representation $\mathbf{h}^{\text{audio}} = [h^{\text{audio}}_{1}, h^{\text{audio}}_{2}, ..., h^{\text{audio}}_{T}] \in \mathbb{R}^{T \times D}$. The predictor converts the transcriptions $\mathbf{y} = [y_{1}, y_{2}, ..., y_{U}] \in \mathbb{R}^{U \times 1}$, where $y_{u}$ is the label token and $U$ is the total number of tokens, into a high-level textual representation $\mathbf{h}^{\text{text}} = [h^\text{{text}}_{1}, h^\text{{text}}_{2}, ..., h^\text{{text}}_{U}] \in \mathbb{R}^{U \times D}$. The joiner, conditioned on both the high-level acoustic $\mathbf{h}^{\text{audio}}$ and textual $\mathbf{h}^\text{{text}}$ sequence, predicts the next token logits $\mathbf{p}^{\text{token}} \in \mathbb{R}^{T \times U \times (V+1)}$, where $V+1$ represents the size of the whole vocabulary and the special blank token $\phi_\text{RNN-T}$. The logits for one step can be formulated as:
\begin{align}
    p^{\text{token}}_{t,u} &= \text{Joiner}\left(h^{\text{audio}}_{t}, h^{\text{text}}_{u}\right),
\end{align}
where $p^{\text{token}}_{t,u} \in \mathbb{R}^{1 \times V}$  and subscript $t$ and $u$ denotes the current acoustic frame index is $t$ and text index is $u$, respectively.

For decoding strategies in Transducer-based KWS, most approaches~\cite{RNNT-KWS_02,RNNT-KWS_03-RNNTByteDance,RNNT-KWS_04,RNNT-KWS_05_CaTT-KWS} adopt ASR decoding techniques to generate hypotheses and match keywords. These algorithms consider the entire search space rather than focusing on the presence of specific keywords, making them sub-optimal for keyword spotting. In~\cite{RNNT-KWS_06_TDT}, the author proposes an autoregressive~(AR) streaming decoding algorithm specifically designed for Transducer-based KWS. This algorithm fixes the predictor input as the keyword during inference and searches within a keyword-specific lattice. The search can be initiated dynamically at any time step, which enables streaming inference. It outputs the most likely frame-level confidence score for the keyword across all paths up to time $t$.  In this paper, we propose a semi-autoregressive~(SAR) decoding algorithm inspired by the AR streaming decoding.

While Transducer-based KWS typically outperforms alternative architectures, it exhibits limitations under challenging conditions. Prior RNN-T-based ASR studies have demonstrated that simple prediction networks often outperform complex ones. In ASR, a single-layer LSTM~\cite{NeMo}, CNN~\cite{kuang22_interspeech}, or stateless embedding-based predictor~\cite{icassp2020-stateless-rnnt} is generally sufficient for modeling contextual dependencies. Following similar design principles and considering resource constraints, KWS systems commonly adopt simple prediction networks. However, as noted in~\cite{RNNT-KWS_03-RNNTByteDance}, keyword-specific data typically contain only the keyword sequence. This characteristic can lead to overfitting in the simple prediction network, as the input transcripts and prediction targets remain fixed during training. When exposed to large amounts of such data, the model may tend to output the keyword under complicated scenarios, resulting in elevated false alarm rates. Unlike hybrid or multi-stage systems~\cite{RNNT-KWS_03-RNNTByteDance,RNNT-KWS_04,RNNT-KWS_05_CaTT-KWS,asru2023-aozhang-u2kws}, we aim to mitigate this overfitting issue by improving the training and decoding framework of the Transducer-based KWS itself.

\section{Methodology}
\subsection{Masked self-distillation training}  
To address the aforementioned overfitting issue, we propose a mask-based self-distilled~(MSD) training method comprising two core components: mask-based training and a self-distillation strategy. Specifically, to reduce the risk of the prediction network overfitting to keyword textual sequences, we apply random zero mask on its output to decrease the heavy reliance on this semantic-modeling module. Then, we require each data sample forward twice and construct teacher-student learning paradigm between the original output and the masked augmented output. The self-distillation operation aims to minimize the converge difficulty introduced by the mask-based learning.

\Cref{fig:overview} illustrates the overall structure of our framework, and the training procedure is detailed in this section. For each utterance in the mini-batch, it will be constructed to two types of training samples:  non-masked and masked ones. Based on a preset mask probability $\gamma_{\text{mask}}$, we apply masking for the masked sample at the token level. Specifically, for the latent textual representation $\mathbf{h}^\text{text}$ derived from the text sequence $\mathbf{y}$, each latent vector $h^\text{{text}}_{u}$ can be masked with probability $\gamma_{\text{mask}}$. This operation is expressed as:
\begin{equation}
    h^{\text{mask}}_{u} = \text{RandomMask}\left(h^{\text{text}}_{u}, \gamma_{\text{mask}}\right).
\end{equation}
This mechanism functions similarly to the dropout strategy~\cite{2014-dropout}, dynamically suppressing portions of the semantic representation from the predictor. It offers two main advantages: (1) For the joint network, stochastic masking introduces variability in the predictor output, even when textual inputs remain constant, thereby enhancing generalization. (2) For the acoustic encoder, although keyword semantics remain fixed, acoustic characteristics such as recording conditions, speaker attributes, or accent are inherently diverse. Masking part of the semantic input encourages the model to focus on these acoustic variations, reducing over-reliance on fixed keyword information and promoting more robust acoustic modeling. The operation on the joint network can be represented as:
\begin{align}
    p^{\text{token}}_{t,u} &= \text{Joiner}\left(h^{\text{audio}}_{t}, h^{\text{text}}_{u}\right),
\end{align}
and
\begin{align}
    p^{\text{mtoken}}_{t,u} &= \text{Joiner}\left(h^{\text{audio}}_{t}, h^{\text{mask}}_{u}\right).
\end{align}
In further, the RNN-T loss for a training pair $(\mathbf{x}, \mathbf{y})$ can be formulated as:
\begin{align}
\mathcal{L}_{\text{RNN-T}}(\mathbf{x},\mathbf{y}) &= -\log p(\mathbf{y} \mid \mathbf{x}) \nonumber \\
&= -\log p(\mathbf{y} \mid \mathbf{h}^{\text{audio}}, \mathbf{h}^{\text{text}}),
\end{align}
and the RNN-T loss with masked predictor output is shown as:
\begin{align}
\mathcal{L}^{\text{mask}}_{\text{RNN-T}}(\mathbf{x},\mathbf{y}) &= -\log p(\mathbf{y} \mid \mathbf{x}) \nonumber \\
&= -\log p(\mathbf{y} \mid \mathbf{h}^{\text{audio}}, \mathbf{h}^{\text{mask}}),
\end{align}
where $\mathbf{h}^{\text{mask}}=[h^\text{{mask}}_{1}, h^\text{{mask}}_{2}, ..., h^\text{{mask}}_{U}] \in \mathbb{R}^{U \times D}$, with $U$ and $D$ representing the sequence length and latent dimension, respectively.

\input{alg}

However, due to the compact nature of KWS models for on-device deployment, our preliminary experiments reveal that Transducers struggle to converge when partial semantic information is removed, reflecting limited learning capacity. To retain discriminative capability and ease training, we introduce a self-distillation mechanism between the original RNN-T logits and those derived from masked predictor outputs. By enforcing consistency between these outputs for each training sample, the model is encouraged to approximate the full distribution even with incomplete predictor input. Accordingly, we propose a masked self-distillation (MSD) strategy to facilitate training with partial or no predictor information, where logits from unmasked inputs serve as soft targets for the masked counterpart. The Kullback–Leibler (KL) divergence is employed to align the two output distributions. For a given utterance, the MSD loss is defined as:
\begin{align}
    \mathcal{L}_{\text{MSD}} = \sum_{t, u} D_{\text{KL}}(p^{\text{token}}_{t,u} \parallel p^{\text{mtoken}}_{t,u}).
\end{align}
The self-teacher-student (STS) learning paradigm helps ensure that the masked output closely resembles the standard transducer output, allowing the model to learn how to produce accurate results without relying on the prediction network. This reduces the overfitting of the predictor to the keyword data. Additionally, the Transducer loss is applied to both the logits corresponding to the unmasked and masked inputs. The total training loss is given by:
\begin{align}
    \mathcal{L} = \mathcal{L}_{\text{RNN-T}} + \lambda_{\text{mask}}\mathcal{L}^{\text{mask}}_{\text{RNN-T}} +  \lambda_{\text{MSD}}\mathcal{L}_{\text{MSD}},
\end{align}
where $\lambda_{\text{mask}}$ and $\lambda_{\text{MSD}}$ are the loss coefficients used to balance the loss contributions.

\subsection{Semi-autoregressive decoding inference}

As discussed in \Cref{sec:2.1}, conventional RNN-Ts perform AR decoding. In contrast, MSD training enables predictor omission during inference, making non-autoregressive (NAR) decoding feasible. As shown in the inference diagram of~\Cref{fig:overview}, our self-distilled Transducer supports both decoding modes: (1) AR decoding, where the model operates as a standard Transducer and employs streaming inference to detect keyword activation. (2) NAR decoding, where the predictor output is pre-masked, allowing only the encoder and joint network to contribute, effectively treating the model as a stateless acoustic model. In both modes, we apply a keyword lattice and the state-of-the-art (SOTA) search algorithm from~\cite{RNNT-KWS_06_TDT} to compute keyword scores at each time step. AR decoding ensures robust performance across general conditions, while NAR decoding proves effective in acoustically challenging scenarios. To leverage the strengths of both, we propose a semi-autoregressive (SAR) decoding strategy tailored for streaming KWS, which aims to leverage both AR and NAR decoding advantages. At time step t, the combined SAR score can be represented as:
\begin{align}
\label{equ:SAR}
\text{Score}^{\text{SAR}}_{t} = \alpha \cdot \text{Score}^{\text{AR}}_{t} \oplus (1-\alpha) \cdot \text{Score}^{\text{NAR}}_{t},
\end{align}
where $\text{Score}^{\text{SAR}}_{t}$, $\text{Score}^{\text{AR}}_{t}$, and $\text{Score}^{\text{NAR}}_{t} \in (0,1)$ denote the activation scores from SAR, AR, and NAR decoding respectively. The operator $\oplus$ denotes the fusion function, and $\alpha$ is a balancing coefficient. For implementation details of SAR decoding, please refer to~\Cref{alg:alg_SAR_algo}.

\section{Experimental Setups}

\subsection{Datasets}

To comprehensively assess the performance and robustness of our model, we evaluate it across three distinct scenarios: (1) EN Fixed, which involves fixed English keywords and serves as a standard KWS benchmark; (2) EN Arbitrary, which evaluate the generalization ability of models when trained solely on general ASR data without specific keyword supervision; and (3) ZH Noisy, which targets fixed Mandarin keyword detection under challenging acoustic conditions to evaluate robustness. The corresponding datasets are summarized as follows:
\begin{itemize}
    \item \textbf{Hey Snips (Snips)}~\cite{snips}: Snips is a widely used KWS benchmark containing positive samples with the keyword ``Hey Snips." However, the official dataset lacks transcriptions for negative samples, which are essential for training RNN-T models. To make full use of the data, we compile all negative samples from the training, development, and test sets into a larger 97-hour negative test dataset.    
    \item \textbf{LibriSpeech and LibriKWS-20}~\cite{RNNT-KWS_06_TDT,LibriSpeech}:    LibriSpeech is a publicly available, high-quality English speech dataset with transcription. We train English RNN-T models on the LibriSpeech 960-hour training set, using them for two purposes. First, these models serve as seed models for training Snips KWS models. Second, they are evaluated for arbitrary KWS performance on the LibriKWS-20. This dataset selects 20 arbitrary keywords from the LibriSpeech test-clean and test-other subsets, and the keywords are listed in \Cref{tab:librikws-20-keywords}.
    \item \textbf{AISHELL-2 and MobvoiHotwords}~\cite{arxiv2018-jiayudu-aishell2,spl2019-jingyonghou-hixiaowen_nihaowenwen}: AISHELL-2 is an open-source Mandarin ASR dataset containing approximately 1000 hours of training data. We utilize AISHELL-2 to train seed RNN-T models for Mandarin KWS experiments. MobvoiHotwords is a Mandarin KWS dataset featuring two keywords, ``nihao wenwen"~(Wenwen) and ``hi xiaowen."~(Xiaowen) Unlike Snips and LibriKWS-20, the MobvoiHotwords recordings including both keyword and non-keyword data, were collected in less controlled environments at various distances from a smart speaker. These recordings include background noise, such as typical household sounds (e.g., music and TV), at varying signal-to-noise (SNR) ratios. As a result, this dataset is more challenging and prone to causing on-device models to overfit to noise, leading to higher false alarm rates.
\end{itemize}

\begin{table}[t]
    \caption{The selected keywords in LibriKWS-20. The keywords for test-clean and test-other two datasets are the same.}
    \label{tab:librikws-20-keywords}
    \centering
    \newcolumntype{S}{>{\small}c}
    \begin{resizebox}{1.0\columnwidth}{!}{
        \begin{tabular}{c c c c c}
        \toprule
        ~& ~&\textbf{Keywords}&~&~ \\
        \midrule
        almost & anything & behind & captain & children \\
        company & continued & country & everything & hardly \\
        himself & husband & moment & morning & necessary \\
        perhaps & silent & something & therefore & together \\
        \bottomrule
        \end{tabular}%
    }
    \end{resizebox}
\end{table}

\subsection{Configuration}
\textbf{Feature extraction.} FBank features with 40 dimensions are extracted using a 25 ms window with a 10 ms shift.Five previous and five subsequent frames are concatenated to create a 440-dimensional encoder input. During training, online speed perturbation~\cite{speed_perturb} is applied with a random ratio selected from \(\{0.9, 1.0, 1.1\}\). Additionally, SpecAugment~\cite{specaug} is employed during training to enhance robustness, applying two random masks on the frequency and time domains with \(f_{\text{max}}=10\) and \(t_{\text{max}}=50\).

\textbf{Model configuration.} The encoder consists of 8-layer Deep Feedforward Sequential Memory Network~(DFSMN)~\cite{DFSMN} blocks with input, hidden, and output sizes of 440, 768, and 320, respectively. The left and right context frames for each DFSMN layer are set to 20 and 8. The stateless predictor, implemented in NeMo~\cite{NeMo}, uses a context size of 2 and an embedding dimension of 320. The joint network combines the 320-dimensional hidden vectors from both the encoder and decoder into a 256-dimensional hidden representation. The final output of the Transducer consists of 70 monophones and 1 additional blank token. The monophones are converted using the grapheme-to-phoneme (G2P) tool~\cite{arxiv-lee-g2p}.

\textbf{Training details.} We employ the AdamW optimizer~\cite{adam-iclr2015, adamw-Ilya-frank-iclr2019} with an initial learning rate of 1e-3 and betas set to (0.9, 0.999). The learning rate is adjusted via the ReduceLROnPlateau scheduler. Training is conducted on 8 NVIDIA V100 GPUs with a mini-batch size constrained by the stricter of 12,288 frames or 64 samples. Loss balancing coefficients are set to $\lambda_{\text{mask}} = 1$ and $\lambda_{\text{MSD}} = 0.003$.

\textbf{Evaluation.} Recall at specific false alarms (FAs) is reported across all datasets. For the LibriKWS-20 test-clean and test-other sets, which contain multiple keywords, we present the macro-recall averaged over all 20 keywords. For the three English test sets (EN Fixed and EN Arbitrary), we set the SAR decoding coefficient to 0.5 as described in \Cref{equ:SAR} and \Cref{alg:alg_SAR_algo}. For the challenging Mandarin Wenwen and Xiaowen datasets (ZH Noisy), we set $\alpha = 0.3$ to give dominance to the NAR score in the SAR decoding process.

\section{Results and Analysis}

\subsection{Optimal mask ratio}
\Cref{tab:mask_prob} presents the KWS performances with different mask probabilities $\gamma_{\text{mask}}$. The model degenerates to a standard RNN-T when $\gamma_{\text{mask}} = 0$. From the table, we observe that the performance generally surpasses that of the standard RNN-T, with the best results achieved when the mask ratio is around 0.35. These results demonstrate the advantages of masked training for RNN-T-based KWS systems and highlight the important role of the predictor. As the mask ratio increases after the optimal value, performance gradually decreases. For all subsequent experiments, we use the optimal mask ratio $\gamma_{\text{mask}} = 0.35$, unless otherwise specified.

\begin{table}[h]
    \caption{The macro-recall of SAR results varies with different mask ratios \(\gamma_{\text{mask}}\) applied to the prediction output on LibriKWS-20. The results are reported at a FA of 4 for all keywords.}
    \label{tab:mask_prob}
    \centering
    \begin{tabular}{c|cc}
    \toprule
        \textbf{$\gamma_{\text{mask}}$} & \textbf{test-clean} & \textbf{test-other} \\
    \midrule
        0.0 & 98.08 & 88.20\\
    \hline
        0.1 & 98.06 & 87.31\\
        0.15 & 98.55 & 93.67\\
        0.2 & 98.70 & 91.52\\
        0.25 & 98.51 & 91.32\\
        0.3 & 98.46 & 92.90\\
        0.35 & \textbf{98.74} & \textbf{93.04}\\
        0.4 & 98.55 & 90.84\\
        0.45 & \textbf{98.74} & 92.02\\
        0.5 & 97.50 & 91.85\\
    \bottomrule
    \end{tabular}
\end{table}

\subsection{The observations on overfitting issues}
\Cref{tab:main} presents the KWS performances of various decoding strategies across several different test scenarios. In this section, we analyze and validate the application scenarios that are prone to overfitting. Comparing the first three rows in~\Cref{tab:main}, we observe that KWS-specific AR (streaming KWS) decoding outperforms ASR-based decoding methods (Greedy Search and Beam Search) on the Snips and LibriKWS-20 datasets, which represent fixed and arbitrary keyword settings, respectively. This trend aligns with findings in~\cite{RNNT-KWS_06_TDT}, reinforcing that streaming decoding is more effective for KWS than ASR-based approaches. These results suggest that when the model is trained on clean keyword or general ASR data, overfitting is not a concern.

However, on noisy keyword data (ZH Noisy), AR decoding shows a significant performance degradation (underlined AR results vs. Greedy and Beam Search). The primary distinction lies in the search space: ASR-based methods operate over the full phoneme space, whereas streaming KWS is confined to a compact, keyword-specific phoneme lattice. Under challenging acoustic conditions, this restricted search, combined with model overfitting, increases the likelihood of false keyword detections, leading to higher false alarm rates and lower recall. In contrast, approaches with broader search spaces exhibit greater robustness. This observation further reinforces the overfitting concern highlighted in~\cite{RNNT-KWS_03-RNNTByteDance}.

\begin{table}[t]
    \caption{Recall at a FAR of 4 among different decoding strategies. For test-clean and test-other containing 20 keywords, we report the macro-recall. \underline{Underline} results indicate the models overfit to noise on Mandarin KWS datasets.}
    \label{tab:main}
    \centering
  \begin{resizebox}{1.0\columnwidth}{!}
  {
    \begin{tabular}{c|ccc|cc}
    \toprule
    
     \multirow{3}*{\textbf{Decoding}} & \multicolumn{5}{c}{\textbf{Recall@\#FA=4}} \\
     \cmidrule(lr){2-6}
     ~ & \textbf{EN Fixed} & \multicolumn{2}{c|}{\textbf{EN Arbitrary}} & \multicolumn{2}{c}{\textbf{ZH Noisy}} \\ 
     \cmidrule(lr){2-2} \cmidrule(lr){3-4}\cmidrule(lr){5-6}
     ~ & \textbf{Snips} & \textbf{test-clean} & \textbf{test-other} & \textbf{Xiaowen} & \textbf{Wenwen} \\
    \midrule
         Greedy Search & 82.13 & 87.65 & 51.69 & 40.28 & 94.56\\
         Beam Search & 89.44 & 87.82 & 54.09 & 83.93 & 95.63\\
         \midrule
         AR~\cite{RNNT-KWS_06_TDT} &  \multirow{2}*{97.31} &  \multirow{2}*{98.55} &  \multirow{2}*{92.17} &  \multirow{2}*{\underline{26.30}} &  \multirow{2}*{\underline{90.44}}\\
         (Streaming KWS) & ~ &~ &~ &~ & \\
         \midrule
         NAR & 97.31 & 94.07 & 83.13 & 95.33 & 96.72\\
         SAR & \textbf{97.35} & \textbf{98.74} & \textbf{93.04} & \textbf{95.68} & \textbf{96.83}\\
    \bottomrule
    \end{tabular}
} \end{resizebox}
\end{table}

\subsection{The effectiveness of SAR decoding strategy}
The previous section demonstrated that fixed-keyword models are prone to overfitting under noisy conditions, and that KWS-specific AR decoding degrades when the model over-predicts the keyword. In this section, we show how the proposed method mitigates this issue and enhances performance.

As shown in the lower part of \Cref{tab:main}, we first compare the proposed mask-based NAR decoding with both ASR-based and AR decoding strategies. Results across the five test sets reveal two key trends. On the EN Fixed and EN Arbitrary datasets, NAR decoding outperforms ASR-based decoding but remains inferior to AR decoding, indicating that the prediction network still contributes useful contextual sementic information in clean scenarios. More importantly, on the ZH Noisy datasets (Xiaowen and Wenwen), where AR decoding suffers from overfitting, NAR decoding achieves substantial gains. Specifically, it significantly outperforms AR decoding (95.33 vs. 25.30; 96.72 vs. 90.44) and even surpasses Beam Search (95.33 vs. 83.93; 96.72 vs. 95.63), which itself exceeds AR performance. These results confirm that our NAR decoding effectively alleviates overfitting in noisy environments.

Furthermore, SAR decoding consistently achieves the best results across all datasets. By fusing strong performance of AR decoding in clean scenarios with robustness to overfitting of NAR decoding, SAR decoding strategy offers a balanced and effective solution. These results demonstrate the generalization and robustness of SAR decoding across both English and Mandarin datasets, under both clean and noisy conditions.


\subsection{The importance of masked self-distillation}

\Cref{tab:ablation_sd} examines the impact of the masked self-distillation (MSD) training. We comprehensively assess this training mechanism across all test datasets and decoding strategies. For both NAR and AR decoding strategies, while MSD training slightly degrades performance when the model without self-distillation is already performing well, it significantly improves performance when the original models are underperforming. Specifically, MSD achieves an average improvement of 14.47\% and 18.03\% for AR and NAR, respectively. The improvement in AR performance suggests that better learning of the masked predictor can help conventional RNN-T models overcome overfitting issues in challenging datasets while maintaining comparable performance on clean test sets. The NAR improvement highlights the difficulty of learning without a predictor output, and MSD effectively assists the encoder and joiner in RNN-T models to better learn to encode audio directly. Evaluation of SAR decoding confirms the universal benefits of MSD. By fully leveraging the advantages of both decoding strategies, MSD achieves superior results across all datasets, with an average improvement of 16.64\%.

\begin{table}[t]
    \caption{The results of the proposed models with and without the masked self-distillation loss \(\mathcal{L}_{\text{MSD}}\) across all 5 test sets using AR, NAR, and SAR decoding strategies are presented. The results are reported at a FA of 4 for all keywords. ``Avg. Imp." denotes the average absolute improvement after applying \(\mathcal{L}_{\text{MSD}}\).}
    \label{tab:ablation_sd}
    \centering
    \begin{resizebox}{1.0\columnwidth}{!}{
    \begin{tabular}{cc|ccccc|c}
    \toprule
         \multirow{3}{*}{\textbf{Decoding}} & \multirow{3}{*}{\textbf{$\mathcal{L}_{\text{MSD}}$}} & \multicolumn{5}{c|}{\textbf{Recall@\#FA=4}} &   \multirow{3}{*}{\textbf{\makecell{Avg. \\ Imp.}}}\\
        \cmidrule{3-7}
         & & \textbf{Snips} & \textbf{test-clean} & \textbf{test-other} & \textbf{Xiaowen} & \textbf{Wenwen} \\
    \midrule

         \multirow{2}{*}{AR} & \ding{56} & 96.56 & \textbf{98.85} & \textbf{93.51} & 4.40 & 9.12 & \\ 
         & \ding{51}  & \textbf{97.31} & 98.55 & 92.17 & \textbf{26.30} & \textbf{60.44}&\textbf{+14.47}\\
    \midrule
         \multirow{2}{*}{NAR} & \ding{56} & 96.48 & 84.05 & 68.97 & 28.49 & \textbf{98.40} & \\ 
            &  \ding{51} & \textbf{97.31} & \textbf{94.07} & \textbf{83.13} & \textbf{95.33} & 96.72 &\textbf{+18.03}\\
    \midrule
         \multirow{2}{*}{SAR} & \ding{56} & 96.32 & 98.55 & 85.11 & 26.91 & 91.53 & \\ 
       & \ding{51} & \textbf{97.35} & \textbf{98.74} & \textbf{93.04} & \textbf{95.68} & \textbf{96.83} &\textbf{+16.64} \\
    \bottomrule
    \end{tabular}
    }\end{resizebox}
\end{table}



\begin{table}[h]
    \caption{Recall values at various low FAs on the Xiaowen dataset, which exhibits the most serious overfitting issue under strict test conditions.}
    \label{tab:overfit_analysis}
    \centering
    \begin{resizebox}{1.0\columnwidth}{!}{
    \begin{tabular}{c|cccccccc}
    \toprule
        \multirow{2}{*}{\textbf{Decoding}} & \multicolumn{8}{c}{\textbf{Recall@\#FAs}} \\
        \cmidrule{2-9}
         & 1 & 2 & 3 & 4 & 5 & 6 & 12 & 24 \\
    \midrule
        AR & 5.90 & 10.81 & 11.53 & 26.30 & 70.55 & 80.13 & 97.72 & 99.49\\
        \midrule
        NAR & 82.10	& 93.81	& 93.99 & 95.33	& 96.23 & 96.36 & 98.37 & 99.46\\
        SAR & \textbf{89.49} & \textbf{94.45} & \textbf{95.22} & \textbf{95.68} & \textbf{96.98} & \textbf{97.18} & \textbf{98.59} & \textbf{99.60} \\
    \bottomrule
    \end{tabular}
    }\end{resizebox}
\end{table}

\subsection{The effectiveness of overfitting suppression}

In \Cref{tab:overfit_analysis}, we report recall values at low FAs on the Xiaowen test set, which shows the worst performance with conventional RNN-T-based KWS, to further analyze the overfitting issue. AR decoding significantly underperforms compared to NAR and SAR at very low FAs ($\text{\#FA} \leq \text{6}$), and eventually achieves comparable results when the FAs increase, such as FAs of 12 and 24 (97.72\% vs. 98.37\% and 99.49\% vs. 99.46\%). In contrast, NAR decoding performs better at very low FAs, highlighting the overfitting suppression capability of our training paradigm when masking the predictor output. SAR consistently delivers the best performance across all decoding strategies, demonstrating its robust stability and generalization under stringent test conditions.


\section{Conclusion}
In this paper, we utilize randomly masked predictor outputs in RNN-T KWS systems to mitigate overfitting issues in challenging scenarios. To enable masked on-device transducer learning, we propose a masked self-distillation (MSD) training paradigm, which restricts the output with or without the prediction network. To leverage the benefits of both training approaches, we introduce the integrated decoding strategy, SAR, which combines the superior performance of AR decoding on normal datasets with the robustness of NAR decoding in scenarios prone to overfitting. The results across three different KWS datasets demonstrate that our MSD training strategy and SAR decoding approach significantly improve the performance of RNN-T-based KWS models.

\newpage
\bibliographystyle{style/IEEEtran}
\bibliography{citations/mybib}

\end{document}

%% file: alg.tex
\begin{algorithm}[!t]
\caption{SAR Transducer Streaming Decoding}
\label{alg:alg_SAR_algo}
\resizebox{0.9\linewidth}{!}{%
\begin{minipage}{\linewidth}


\KwIn{%
    Phoneme/blank token posterior matrix $\mathbf{p}^{\text{token}} = [p_1,p_2,\dots,p_T] \in \mathbb{R}^{T\times U\times V}$,\\
Masked token posterior matrix 
$\mathbf{p}^{\text{mtoken}} \in \mathbb{R}^{T\times U\times V}$,
with entries 
$[p^{\text{mask}}_1, p^{\text{mask}}_2, \dots, p^{\text{mask}}_T]$,\\
    Keyword phoneme sequence $\mathbf{y}=[y_1,y_2,\dots,y_U]$,\\
    Fusion coefficient: $\alpha$ \\
    Bonus score: $S_{\mathrm{Bonus}}$,\\
    Timeout: $T_{\mathrm{out}}$.
}
\KwOut{SAR activation scores $\mathbf{S}[1:T]$}
\BlankLine

\textbf{Initialization:}\\
1) Insert RNN-T blank token: \\
\Indp
$\mathbf{y} = [\phi_{\text{RNN-T}}, y_1, y_2, \dots, y_U]$. \\
\Indm
2) Initialize SAR scores: $\mathbf{S}[1:T] = \{0\}$. \\
3) Boundary conditions: \\
\Indp
$\phi(0, u) = \phi^\text{mask}(0, u) = 0$, \\
$\delta(0, u) = \delta^\text{mask}(0, u) = 1$, for $0 \le u \le U$. \\
\Indm
4) Set $t = 1$. 

\While{$t \le T$}{
    \If{$u = 0$}{
        $\delta(t, u) = \delta^{\text{mask}}(t, u) = 1$ \tcp*{Initialize new path at time $t$}
    }
    
    \For{$u = 1$ \KwTo $U$}{
        $\delta(t, u) = \max\left\{
        \begin{aligned}
            &\delta(t, u{-}1) \cdot p_{t, u{-}1}(\mathbf{y}_{u}), \\
            &\delta(t{-}1, u) \cdot p_{t{-}1, u}(\phi_{\text{RNN-T}})
        \end{aligned}
        \right.$ \;
        \BlankLine
        \BlankLine
        $\delta^{\text{mask}}(t, u) = \max\left\{
        \begin{aligned}
            &\delta^{\text{mask}}(t, u{-}1) \cdot p^{\text{mask}}_{t, u{-}1}(\mathbf{y}_{u}), \\
            &\delta^{\text{mask}}(t{-}1, u) \cdot p^{\text{mask}}_{t{-}1, u}(\phi_{\text{RNN-T}})
        \end{aligned}
        \right.$ \;
    }

    \tcp{Compute SAR decoding score at time $t$}
    $\mathbf{S}[t] = \Bigl(
        \alpha \cdot \delta(t, U) \cdot \phi(t, U) \; +\;
        (1-\alpha) \cdot \delta^\text{mask}(t, U) \cdot \phi^\text{mask}(t, U)
    \Bigr)$ \;

    \BlankLine
    \tcp{Compute path length $\ell(t)$ for normalization}
    $\ell(t) = \text{RecordPathLength}(\mathbf{S}[t])$ \;
    
    \BlankLine
    \If{$\ell(t) > T_{\mathrm{out}}$}{
        \tcp{Prune over-length paths}
        $\mathbf{S}[t] = 0$ \;
    }
    
    $\mathbf{S}[t] = \text{pow}\Bigl(S_{\mathrm{Bonus}} \cdot \mathbf{S}[t],\, 1 / \ell(t)\Bigr)$ \;
    $t = t + 1$ \;
}
\Return $\mathbf{S}[1:T]$
\end{minipage}
}
\end{algorithm}